\def   \ni {\noindent}

\def   \ssk {\vskip  5truept}

\def   \bsk {\vskip 15truept}

\def   \newline {\hfil\break}

\def\ga{\gamma}

\def\la{\lambda}

\def\si{\sigma}

\documentstyle[epsfig]{article}
\begin{document}

%
% A&A definitions
\def\la{\mathrel{\mathchoice {\vcenter{\offinterlineskip\halign{\hfil
$\displaystyle##$\hfil\cr<\cr\sim\cr}}}
{\vcenter{\offinterlineskip\halign{\hfil$\textstyle##$\hfil\cr
<\cr\sim\cr}}}
{\vcenter{\offinterlineskip\halign{\hfil$\scriptstyle##$\hfil\cr
<\cr\sim\cr}}}
{\vcenter{\offinterlineskip\halign{\hfil$\scriptscriptstyle##$\hfil\cr
<\cr\sim\cr}}}}}
\def\ga{\mathrel{\mathchoice {\vcenter{\offinterlineskip\halign{\hfil
$\displaystyle##$\hfil\cr>\cr\sim\cr}}}
{\vcenter{\offinterlineskip\halign{\hfil$\textstyle##$\hfil\cr
>\cr\sim\cr}}}
{\vcenter{\offinterlineskip\halign{\hfil$\scriptstyle##$\hfil\cr
>\cr\sim\cr}}}
{\vcenter{\offinterlineskip\halign{\hfil$\scriptscriptstyle##$\hfil\cr
>\cr\sim\cr}}}}}
\def\degr{\hbox{$^\circ$}}
\def\arcmin{\hbox{$^\prime$}}
\def\arcsec{\hbox{$^{\prime\prime}$}}

\hsize 5truein
\vsize 8truein
\font\abstract=cmr8
\font\keywords=cmr8
\font\caption=cmr8
\font\references=cmr8
\font\text=cmr10
\font\affiliation=cmssi10
\font\author=cmss10
\font\mc=cmss8
\font\title=cmssbx10 scaled\magstep2
\font\alcit=cmti7 scaled\magstephalf
\font\alcin=cmr6 
\font\ita=cmti8
\font\mma=cmr8
\def\ref{\par\noindent\hangindent 15pt}
\null
%\vskip 3.0truecm
%\baselineskip = 12pt

% beginning of font "title"

\title{\ni
 CMB and LSS Power Spectra From Local Cosmic String Seeded Structure Formation
}                                               

% beginning of font "author and affiliation"
\bsk \bsk
\author{\ni Carlo R. Contaldi$^{1}$, Mark Hindmarsh$^2$ and 
Jo\~ao Magueijo$^1$}                                                       
\bsk
\affiliation{\ni $^1$ Theoretical Physics, The Blackett Laboratory,
Imperial College, Prince Consort Road, London, SW7 2BZ, U.K.\\ 
$^2$ Centre for Theoretical Physics, University of Sussex, 
Brighton BN1 9QJ, U.K. 
}                                                
\bsk
\baselineskip = 12pt

% beginning of font "abstract and keywords"
\abstract{ \ni

We evaluate the two point functions of the stress energy from the
largest string simulations 
carried out so far. The two point functions are used to calculate
the cosmic microwave 
background (CMB) and cold dark matter (CDM) power spectra from local
cosmic string models for structure formation. We find that our spectra
differ significantly from those previously calculated for both
global and local
defects. We find a higher Doppler peak at $l=400-600$ and a less severe
bias problem than for global defects. Spectra were obtained for a
variety of network energy-decay mechanisms.

}                                                    
\bsk
\baselineskip = 12pt

\bsk
\baselineskip = 12pt

% beginning of font "text"

\text
\ni 1. INTRODUCTION
\ssk

The next few years will see a great increase in the accuracy of the
data mapping the CMB temperature and large scale structure (LSS) due to
satellite, balloon and ground based experiments. Similarly accurate
theoretical predictions from inflation and defect models will be required
for direct comparison with the new data. While calculations of structure
formation from 
inflationary models are relatively straightforward, defects, being
highly non-linear objects and active sources of perturbations, pose a
more challenging problem.

To this end we report on what we believe to be the most complete numerical
treatment of local cosmic strings to date. The method we use is
based on that 
described in \cite{pst} where it was shown how to use the two point 
functions (known as the unequal time correlators, or UETCs) measured
form defect
simulations to calculate the CMB and LSS power spectra. The method was
applied to global defects and the results were discouraging for defect
models.

In this work we aimed to verify these results in local theories where
the calculations are complicated somewhat by considerations on energy
conservation and have only been permitted following recent advances
in the computer technology at our disposal. 
The method we used is
considerably different from that used in previous local string
calculations based on simulations \cite{steb} and we believe that the
UETC method is justified by causality and scaling arguments in
dramatically extending the dynamical range of defect simulations. We
also find that some of the our UETCs are approximated quite well by
those based on an analytical model \cite{vinc,abr} but others differ
quite significantly and in \cite{chm} we show how this model can be
elaborated to explain the features observed in the simulations.

\bsk
\ni 2. THE CORRELATORS
\ssk

The perturbations arising form scalar, vector and tensor (SVT)
modes are completely decoupled and evolve separately. To calculate
the CMB and LSS power spectra efficiently, it is therefore
convenient to decompose the source of the perturbations into
irreducible components under rotations. 

Let $\Theta_{\mu\nu}({\bf x})$ be the defect stress-energy tensor.  We
may Fourier analyze it
\begin{equation}
\Theta_{\mu\nu}({\bf x})={\int d^3k}\Theta_{\mu\nu}({\bf k})
e^{i{\bf k}\cdot {\bf x}}
\end{equation}
and decompose its Fourier components as:
\begin{eqnarray}
\Theta_{00}&=&\rho^d\\
\Theta_{0i}&=&i{\hat k_i}v^d+\omega^d_i\\
\Theta_{ij}&=&p^d\delta_{ij}+
{\left({\hat k_i}{\hat k_j}-{1\over 3}\delta_{ij}\right)}
\Pi^S+\nonumber\\
&&i{\left({\hat k_i}\Pi_j^V+{\hat k_j}\Pi_i^V\right)}
+\Pi_{ij}^T
\end{eqnarray}
with ${\hat k^i}\omega^d_i=0$, ${\hat k^i}\Pi^V_i=0$,
${\hat k^i}\Pi_{ij}^T=0$, and $\Pi^{Ti}_{i}=0$.
The variables $\{\rho^d, v^d, p^d, \Pi^S\}$
are the scalars, $\{ \omega^d_i, \Pi^V_i\}$ the vectors,
and $\Pi^T_{ij}$ the tensors.

The unequal time correlators of these components will completely
specify the evolution of the network's stress-energy.  All
correlators between modes at $({\bf k},\eta)$ and $({\bf
k}',\eta')$ will be proportional to $\delta({\bf k}-{\bf k}')$ due
to translational invariance. We shall drop this factor in all
formulae. The correlators can also be functions of $k$ alone, due
to isotropy.  Since conjugation corresponds to ${\bf k}\rightarrow
-{\bf k}$ isotropy implies that the correlators must be
real. Because of incoherence the correlators will be generic
functions of $\eta$ and $\eta'$.

On top of this, the form of the S+V+T decomposition fixes further
the form of the correlators. One can always write down the most
general form of a correlator, and then contract the result with
${\hat k_i}$ or $\delta_{ij}$, wherever appropriate, to obtain
further conditions. Proceeding in this way we can show that cross
correlators involving components of different type (S, V, or T)
must be zero. Furthermore one has that the only non-zero
correlators are the 10 scalar correlator functions
\begin{equation}
\begin{array} {cccc} 
f^{\rho^d\rho^d}& f^{\rho^d v^d}& f^{\rho^d p^d}& f^{\rho^d \Pi^S}\\
\cdots&f^{v^d v^d}&f^{v^dp^d} &f^{v^d \Pi^S}\\ 
\cdots&\cdots &f^{p^d p^d}&f^{p^d \Pi^S}\\
\cdots&\cdots &\cdots &f^{\Pi^S\Pi^S}\\
\end{array}
\end{equation}
the 3 vector correlators:
\begin{equation}
\begin{array} {cc}
f^{\omega^d\omega^d}&f^{\omega^d\Pi^V}\\
\cdots& f^{\Pi^V\Pi^V}
\end{array}
\end{equation}
and the single tensor correlator function $f^{\Pi^T\Pi^T}$.

In general these functions are functions of $(k,\eta,\eta')$,
and this is indeed the case during the matter radiation transition.
However well into the matter and radiation epochs there is scaling,
and these functions may be written as:
\begin{equation}
f^{XY}(k,\eta,\eta')={F^{XY}(x,x')\over \sqrt{\eta\eta '}}
\end{equation}
where $XY$ represents any pair of superscripts considered above,
and $x=k\eta$ and $x'=k\eta'$.
The purpose of this work is the measurement of the 14 functions
of two variables $F^{XY}(x,x')$.

\bsk
\ni

\bsk
\ni 3. NUMERICAL DETERMINATION OF THE CORRELATORS
\ssk

Local strings have an extra complication over global defects,
which stems from the fact that we are unable to simulate the
underlying field theory.  Instead, we approximate the true
dynamics with line-like relativistic strings using the Nambu
equations of motion. We used a previously developed implementation
\cite{col,smi} of this algorithm which simulates the network of
strings neglecting damping effects due to the
expansion of the Universe. The velocities of the strings segments
in the network can be constrained to be integers due to the gauge
conditions used to discretise the equations of motion and this
enables the code to obtain very high accuracy with extremely fast
computation.

To simulate the extraction of energy from the system due to the
decay of the string loops into gravitaional radiation, loops of a
minimum size are excised from the simulation at each
timestep. This also ensures that the network scales with respect
to the conformal time $\eta$ which enables us to extend the
dynamical range of the resulting correlation functions beyond the
limited range covered in the simulation.

We performed simulations with box sizes ranging from $128^3$ to
$600^3$, with a cut-off on the loop size of two links. Realisation
averages were carried out with $256^3$ boxes once it was
determined that the general form of the correlators scaled very
accurately with box size.To evaluate the
UETCs from the simulations we selected times in the range $0.1
N<t<N/4$, where $N$ is the box size, when we were sure that the
string network was scaling, and when boundary effects are still
excluded by causality. From this we obtained the time evolution of the
network's stress-energy tensor $\Theta_{\mu\nu}$ at each point on
the lattice using
\begin{equation}
\Theta_{\mu\nu}({\bf{x}})={\mu\int d\si
(\dot{X}^{\mu}\dot{X}^{\nu}-\acute{X}^{\mu}\acute{X}^{\nu})\delta^3({\bf{x}}-\bf{X}(\si,\eta))}  
\end{equation}
where $X(\si,\eta)$ is the spacetime coordinate of the string with
$\si$ a parameter along the string.
SVT decomposition was carried out on the 10 independent components
of the Fourier transform of the stress-energy tensor at each
timestep. This resulted in all the SVT components being computed
directly from the simulation without making assumptions on energy
conservation and on the details of energy dissipation from the
string network. The drawback of obtaining all the components
directly in such a manner is that the process becomes
computationally intensive for even modestly sized simulations
e.g. $256^3$.
By cross-correlating the decomposed stress-energy components from
a central time with those from all the stored timesteps the 14
independent UETC's $f^{XY}(k,\eta,\eta')$ were computed.

Standard codes for structure formation \cite{cmbfast} require the
square root of 
coherent correlators for sourcing the inhomogeneous differential
equations but this is not straightforward in general defect
scenarios because the correlators are incoherent and therefore do
not factorise.  To overcome this, following \cite{pst}, we expand
the matrix of scaling functions onto a basis of coherent functions
$v^{(i)}_{\alpha\beta}(k\eta)$ 
by diagonalising the matrix of correlators
\begin{equation}\label{eig}
c_{\mu\nu,\alpha\beta}(k\eta,k\eta ')={\sum_i}\lambda^{(i)}
v^{(i)}_{\mu\nu}(k\eta)v^{(i)}_{\alpha\beta}(k\eta')
\end{equation}
where $\lambda^{(i)}$ are eigenvalues. Each coherent, weighted
eigenmode is then fed into the standard codes and the total $C_l$
and $P(k)$ are then the convergent sums of the resulting
components.

\ssk
\ni

\bsk
\ni 4. RESULTS
\ssk

The most striking difference between our work on local strings and
previous work on global defects is that we find the $\Theta_{00}$
term comes to dominate over all the other components and thus
vector and tensor components contribute much less significantly in
our results. The string anisotropic stresses are in the predicted
\cite{tps} ratios $|\Theta^S|^2:|\Theta^V|^2:|\Theta^T|^2$ of
$3:2:4$, as $k\tau\rightarrow 0$, however $|\Theta_{00}|^2\gg |
\Theta^S|^2$. Also we find that the peak in the energy density power
spectra occurs at $k\tau\approx 20$ which is well inside the
horizon.  In general this leads to a higher Doppler peak than
observed in \cite{pst,abr}. 

For a full treatment we must also include in the calculations the
background fluid into which the network of strings dumps
energy. One of the greatest unknowns in the dynamics of cosmic
strings is the channel through which this energy is discarded and
this is quantified by assuming an equation of state for the extra
fluid,of the form $p^X=w^X\Theta^{X}_{00}$, where $w^X$ is varied in
the range $0<w^X<1/3$ so as to account for the non-relativistic case
through to the relativistic case.  

In Fig.1 we plot $\surd[\ell(\ell+1)C_\ell/2\pi]$,  
setting the Hubble constant to
$H_0=50$ Km sec$^{-1}$ Mpc$^{-1}$, the baryon fraction to
$\Omega_b=0.05$, and  assuming a flat geometry, no cosmological
constant, 3 massless neutrinos, standard recombination,
and cold dark matter. 
The most interesting feature is 
the presence of a reasonably high Doppler peak at $\ell=400-600$, 
following a pronouncedly tilted large angle plateau.
This feature sets local strings apart from global defects.
It puts them in a better shape to face the current data.

\begin{figure}
\centerline{\psfig{file=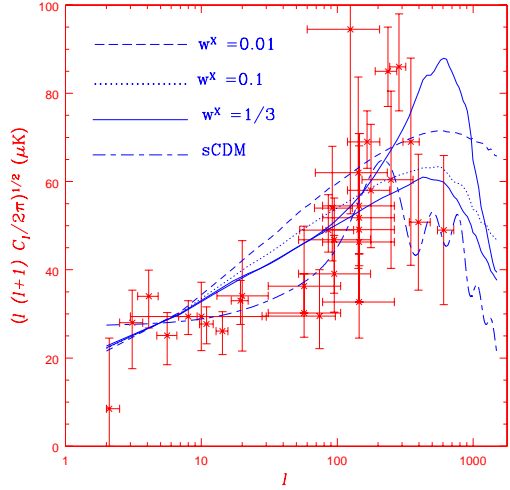,width=7 cm,angle=0}}
\caption{Fig.1 The CMB power spectra predicted by cosmic strings decaying
into loop and radiation fluids with $w^X=1/3, 0.1, 0.01, 0$. 
We have plotted $(\ell(\ell+1)C_\ell/2\pi)^{1/2}$ in $\mu K$,
and superposed several experimental points.
fluid.}
\label{fig1}
\end{figure}
\begin{figure}
\centerline{\psfig{file=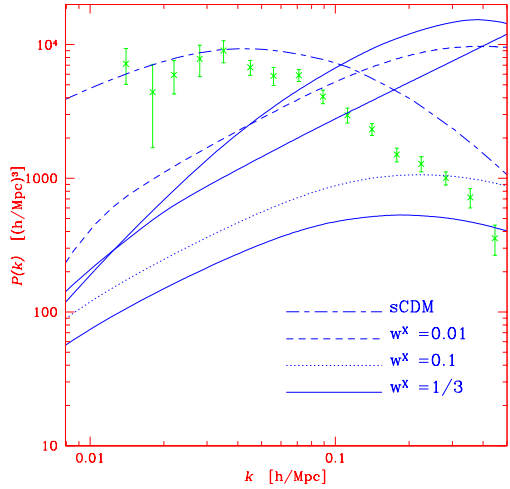,width=7 cm,angle=0}}
\caption{Fig.2 The power spectrum in CDM fluctuations for
cosmic strings, with $w^X=0.01,0.1,1/3$. We plotted
also the standard CDM scenario prediction and points inferred
by Peacock and Dodds from galaxy surveys.}
\label{fig2}
\end{figure}
In Fig.2 we plotted the CDM power spectrum $P(k)$ 
together with experimental points as in \cite{pdodds}.
The normalization has been fixed by COBE data points.
We see that the peak of the spectrum is always at smaller scales 
than standard CDM predictions, or observations. 
The CDM rms fluctuation in 8 $h^{-1}$Mpc spheres is $\sigma_8=.42,
.61, 1.8$ for $w^X=1/3,0.1,0.01$. Hence relativistic decay products
match well the observed $\sigma_8\approx 0.5$. On the other hand
in 100 $h^{-1}$Mpc spheres one requires bias $b_{100}=
\sigma_{100}^{data}/\sigma_{100}=4.9, 3.7, 1.6$
to match observations.  

In this work we have calculated the CMB and LSS power spectra
arising from local cosmic string scenarios using the largest string
simulations to date. The spectra account for a variety of energy
decay mechanisms and the result we want to highlight is the
significant difference between structure formation processes arising
from global and local defects which show that the 100 $h^{-1}$ Mp$c^{-1}$
bias problem and low Doppler peaks are not as generic effects as was
previously thought.

\ssk
\ni

\bsk
\baselineskip = 12pt
{\abstract \ni ACKNOWLEDGMENTS
We tank A. Albrecht, R. Battye and P. Ferreira for useful
discussions. Special thanks are due to 
U-L. Pen, U. Seljak, and N. Turok for giving us their global defect UETCs, 
and to P. Ferreira for letting us use and modify his string code
 This work
was performed on COSMOS, the Origin 2000 supercomputer owned
by the UK-CCC and supported by HEFCE and PPARC.
We acknowledge financial support from the Beit Foundation (CRC), 
PPARC (MH), and the Royal Society (JM).

}

\bsk
\baselineskip = 12pt

% beginning of font "references"
 \ni REFERENCES

\end{document}